\documentclass[fleqn,usenatbib]{mnras}
\usepackage{newtxtext,newtxmath}
\usepackage[T1]{fontenc}
\usepackage{ae,aecompl}

\usepackage{graphicx}	% Including figure files
\usepackage{amsmath}	% Advanced maths commands
\usepackage{amssymb}	% Extra maths symbols

\title[]{The pre-SN activity as a possible explanation of the peculiar properties of Type IIP Supernova 2009kf}

\author[R. Ouchi \& K. Maeda]{
Ryoma Ouchi,$^{1}$\thanks{E-mail: ouchi@kusastro.kyoto-u.ac.jp}
Keiichi Maeda,$^{1}$
\\
$^{1}$Department of Astronomy, Kyoto University, Kitashirakawa-Oiwake-cho, Sakyo-ku, Kyoto 606-8502, Japan \\
}

\date{Accepted XXX. Received YYY; in original form ZZZ}

\pubyear{2020}

\begin{document}
\label{firstpage}
\pagerange{\pageref{firstpage}--\pageref{lastpage}}
\maketitle

\begin{abstract}
  SN 2009kf is an exceptionally bright Type IIP Supernova (SN IIP) discovered by the Pan-STARRS 1 survey. The $V$-band magnitude at the plateau phase is $M_{V} = -18.4$ mag, which is much brighter than typical SNe IIP. We propose that its unusual properties can be naturally explained, if we assume that there was an super-Eddington energy injection into the envelope in the last few years of the evolution before the SN explosion. Using a progenitor model with such an pre-SN energy injection, we can fit the observational data of SN 2009kf with the reasonable explosion energy of $E_{\mathrm{exp}} = 2.8 \times 10^{51}$ erg and the $^{56}$Ni mass of $0.25 M_{\odot}$. Specifically, we injected the energy into the envelope at a constant rate of $3.0 \times 10^{39}$ erg s$^{-1}$ in the last 3.0 years of evolution before the core collapse. We propose that some of the unusually bright SNe IIP might result from the pre-SN energy injection to the envelope.
  %In our model, the dense shell is produced at the outer edge of the envelope as a result of pre-SN energy injection. The early bump seen in the bolometric light curve of SN 2009kf can naturally be explained as a result of the interaction of the shock with the shell. Thus, our work indicates that the existence of the dense circumstellar medium around some supernova might be originated from the pre-SN energy injection due to e.g. gravity waves, unstable nuclear burning.
\end{abstract}


%%%%%%%%%%%%%%%%%%%%%%%%%%%%%%%%%%%%%%%%%%%%%%%%%%

%%%%%%%%%%%%%%%%% BODY OF PAPER %%%%%%%%%%%%%%%%%%

\section{Introduction}

A core collapse supernova is the explosion of a massive star ($\gtrsim 8 M_{\odot}$), which is triggered by the collapse of the iron or oxygen-neon-magnesium core. Especially, type II supernovae (SNe II) are defined by the presence of the hydrogen lines in their spectra. SNe II which have a plateau in their light curves are defined as SNe IIP, while those which have linearly declining light curves are defined as SNe IIL \citep{1997ARA&A..35..309F}.

The SN IIP is the most common class among the core collapse SNe, occupying $\sim 50$ \% of all the observed core collapse SNe \citep{2011MNRAS.412.1441L}. SNe IIP are diverse both in their luminosities and plateau lengths, and they have the typical plateau magnitude of $M_V \sim -16$ -- $-17$ mag \citep{2003ApJ...582..905H, 2014ApJ...786...67A, 2016MNRAS.459.3939V}. The explosion energy of SNe IIP is in the range of $\sim (0.5$ -- $4.0) \times 10^{51}$ erg, which is supposedly aided by the energy deposition by neutrinos \citep{2009ApJ...703.2205K}. The range of the mass of synthesized $^{56}$Ni, on the other hand, is broad, i.e. $0.005$ -- $0.28 M_{\odot}$ \citep{2017ApJ...841..127M}.

It has been well established that SNe IIP are the explosions of red supergiants based on the light curve models \citep{1977ApJS...33..515F, 2003MNRAS.338..939E, 2011ApJ...729...61B}. Also, the red supergiant progenitors have been confirmed by the direct detection of the progenitors in the pre-SN images for a good number of SNe IIP, with their initial masses lying in the range of $8$ -- $17 M_{\odot}$ \citep{2009ARA&A..47...63S, 2015PASA...32...16S}. 

% Moreover, SNe IIP have been demonstrated as a good distance indicator \citep{2002ApJ...566L..63H, 2010ApJ...715..833O}. It is thus important to understand the diversity of SNe IIP.

SN IIP 2009kf was discovered on 2009 June 10.9 UT by the Pan-STARRS 1 (PS1) survey. Galaxy Evolution Explorer (GALEX) also detected the same object in the near $UV$ band. The redshift was estimated as $z=0.182 \pm 0.002$. The mid-plateau magnitude of SN2009kf was $M_V \approx -18.4$ mag. Thus, it is 1.5-2 mag brighter than typical SNe IIP. Furthermore, the hydrogen lines show that it had the expansion velocities of 9000 km s$^{-1}$ at 61 days after the explosion, which is unusually high for SNe IIP \citep{2010ApJ...717L..52B}.

\citet{2010ApJ...723L..89U} conducted a hydrodynamical modeling of SN2009kf and proposed an extremely high explosion energy of $2.2 \times 10^{52}$ erg, and a very large ejecta mass of $28.1 M_{\odot}$, together with a relatively large progenitor radius of $2 \times 10^3 R_{\odot}$, in order to fit the observational data. They claimed that the high explosion energy might be achieved by the aid of the rapid disk accretion onto a black hole as a result of a stellar merger.

On the other hand, \citet{2011MNRAS.415..199M} proposed that the interaction of the SN ejecta with the dense circumstellar medium (CSM) around the progenitor can also explain the observational data. They derived the mass-loss rate of $\dot{M} = 10^{-2} M_{\odot}$ yr$^{-1}$, which extends to $2 \times 10^{15}$ cm. Note, however, that they attach the power-law CSM density structure above the stellar surface being agnostic of how it was created.

Actually, observational evidence has been accumulating showing that massive stars may experience the enhanced mass loss just prior to the core collapse, which then constructs a dense CSM around the progenitor star \citep[see ][and references therein]{2014ARA&A..52..487S}. The spectra characterized by emission lines from highly-ionized ions (the so-called flash spectra) which are taken immediately after the explosion imply that at least a fraction of SN IIP progenitors show the elevated pre-SN mass loss \citep{2017NatPh..13..510Y}. \citet{2016ApJ...818....3K} have found that at least $18 \%$ of SNe II have such flash-ionized spectra by analyzing their samples. Moreover, \citet{2017ApJ...838...28M, 2018ApJ...858...15M} and \citet{2018NatAs.tmp..122F} modeled the early-time light curves of SNe II and found that they are better fit with the dense CSM. Thus, the enhanced pre-SN mass loss seems to be a common property among SNe II.

Although the physical mechanism behind this phenomenon is not yet understood, the energy deposition into the envelope just before the core collapse has been proposed as a possible cause \citep{2010MNRAS.405.2113D, 2014ARA&A..52..487S}.  For example, a fraction of the gravity waves generated from the convective core may tunnel towards the envelope, and deposit energy there \citep{2012MNRAS.423L..92Q, 2017MNRAS.470.1642F, 2018MNRAS.476.1853F}. The energy deposition rate by this process is expected to exceed the Eddington luminosity only in the last few years before the core collapse. Therefore, it naturally explains the finely tuned timing of the mass loss close to the core collapse \citep{2014ApJ...780...96S}. %The hydrodynamic instabilities driven by convection in the core might help to inject energy to the envelope at the advanced burning stages \citep{2011ApJ...733...78A, 2014ApJ...785...82S}.
Explosive shell burning instabilities in the advanced stages might also be the candidate for the additional energy injection \citep{2011ApJ...733...78A, 2014ApJ...785...82S}. As another possibility, \citet{2012ApJ...752L...2C} proposed a common envelope interaction as a possible cause of the pre-SN mass loss, where a companion deposits its orbital energy into the primary's envelope. 

Several works have been done to investigate how such an energy injection affects the structure of stellar envelope \citep{2012MNRAS.423L..92Q, 2017MNRAS.470.1642F, 2018MNRAS.476.1853F}. It has been shown that the super-Eddington energy injection into the envelope causes the dynamical expansion of the envelope, rather than creating a moderate mass-loss wind above the stellar surface \citep{2014MNRAS.445.2492M, 2019ApJ...877...92O}.
%Moreover, it produces the dense shell above the inflating bubble \citep{2016MNRAS.458.1214Q}.
\citet{2019ApJ...877...92O} has shown that the supernova from a progenitor which experienced the super-Eddington energy injection in the last few years of evolution will be an order of magnitude brighter than typical SNe IIP. Although such models can not explain typical SNe IIP, they may be applicable to peculiar SNe IIP.

In this paper, we propose that the model with a super-Eddington energy deposition, which takes place in the last few years of stellar evolution before the core collapse, can actually explain the observational properties of SN2009kf without assuming unusually high explosion energy. This paper is organized as follows; in \S \ref{method}, we describe the set up of the numerical models. In \S \ref{density_progenitor}, \ref{LC}, \ref{sec:Tph} and \ref{vph}, we compare our model with the observational dataset. In \S \ref{discussion}, we discuss our results in relation to the previous works.

\section{Methods} \label{method}

\begin{table}%[htbp]
%\renewcommand{\thetable}{\arabic{table}}
  %  \centering
\caption{Summary of progenitor and explosion properties of all the models.} \label{tab:decimal}
\hspace{-0.6cm}
\scalebox{0.8}{
\begin{tabular}{lccccccc}
%\tablewidth{0pt}
\hline
\hline
Model          &  $M_{\mathrm{ZAMS}}$ &  $M_{\mathrm{final}}$ & $R$ \footnotemark[1]   &   $E_{\mathrm{exp}}$   & $^{56}$Ni     & $L_{\mathrm{dep}}$   \\
               & ($M_{\odot}$)       &  ($M_{\odot}$)    & ($R_{\odot}$) &  (10$^{51}$ erg)               & ($M_{\odot}$) &  (10$^{39}$ erg s$^{-1}$)  \\
\hline
\hline
%\decimals
inject\_M12           &  12             &    10.90    &     6125      &      2.8       & 0.25   &   3.0   \\
inject\_M18           &  18             &    13.52    &     3981      &      2.8       & 0.25   &   2.0   \\
\hline
No\_inject\_M12\_1    & 12              &    10.90      &      736      &      0.5       & 0.03   &   0.0   \\
No\_inject\_M12\_2    & 12              &    10.90      &      736      &      1.0       & 0.03   &   0.0   \\
No\_inject\_M12\_3    & 12              &    10.90      &      736      &      4.0       & 0.25   &   0.0   \\
No\_inject\_M18\_1    & 18              &    13.52      &     1180      &      0.5       & 0.03   &   0.0   \\
No\_inject\_M18\_2    & 18              &    13.52      &     1180      &      1.0       & 0.03   &   0.0   \\
No\_inject\_M18\_3    & 18              &    13.52      &     1180      &      4.0       & 0.25   &   0.0   \\
No\_inject\_M20\_1    & 20              &    14.11      &     1373      &      0.5       & 0.03   &   0.0   \\
No\_inject\_M20\_2    & 20              &    14.11      &     1373      &      1.0       & 0.03   &   0.0   \\
No\_inject\_M20\_3    & 20              &    14.11      &     1373      &      4.0       & 0.25   &   0.0   \\
\hline
%\multicolumn{5}{c}{NOTE. - Two decimal aligned columns}
\end{tabular}
}
%\vspace{-0.3cm}
\label{table}
\end{table}
\footnotetext[1]{The radius of the outer most cell.}     

For the calculation of stellar evolution, we use the one-dimensional stellar evolution code, Modules for Experiments in Stellar Astrophysics of version 10398 \citep[MESA][]{2011ApJS..192....3P, 2013ApJS..208....4P, 2015ApJS..220...15P, 2018ApJS..234...34P}. The set up of the stellar evolution calculation closely follows \citet{2019ApJ...877...92O}. Here, we briefly summarize the key points. We assume the initial metalicity of Z=0.02 and set the mixing length parameter to be $\alpha = 1.9$.

We calculate several progenitor models, including the model with the pre-SN energy injection and those without. As for the model with the pre-SN energy injection (inject\_M12), we firstly evolve a $12M_{\odot}$ non-rotating star from a pre-main sequence to 3 years before the core collapse, without the energy injection. Then, we start injecting energy into the envelope, with the hydrodynamic mode of MESA on, and evolve the model in the last 3 years until the time of the core collapse. The timescale of 3 years is chosen to represent the timescale of the pre-SN activities suggested by several observational studies \citep{2017NatPh..13..510Y}. This timescale is also consistent with the timescale on which the energy injection in the envelope by gravity waves is predicted to be significant \citep{2014ApJ...780...96S, 2017MNRAS.470.1642F}.

Here, we inject energy assuming the gaussian function. For the energy injection rate per mass, we use the following form:
\begin{eqnarray}
\epsilon_{\mathrm{inject}} = \frac{L_{\mathrm{dep}}}{\sigma \sqrt{2 \pi}} \exp \left(-\frac{(r - R_{\mathrm{dep}})^2}{2 \sigma^2} \right) \frac{dr}{dm},
\end{eqnarray}
where $dr$, $dm$ are the width of a numerical cell in the radial and mass coordinates, respectively. We assume $R_{\mathrm{dep}}=600 R_{\odot}$, and $\sigma = 10.0 R_{\odot} $. The stellar radius of the progenitor with the initial mass of $12M_{\odot}$ before the energy injection is $\sim  741R_{\odot}$. So, the present model represents a situation in which the energy deposition takes place close to the stellar surface. For the energy injection rate integrated in mass coordinate, we adopt a constant value of $L_{\mathrm{dep}}= 3.0 \times 10^{39}$ erg s$^{-1}$. Note that the Eddington luminosity of the progenitor is $\approx 5 \times 10^{37}$ erg s$^{-1}$, and thus, the injection rate is more than one order of magnitude higher than the Eddington luminosity.
 
Once the model described above is evolved dynamically to the time of the core-collapse, this is used as an input model for the radiation hydrodynamic simulations of the SN explosions. For this purpose, we use both the codes MESA and STELLA \citep{1998ApJ...496..454B, 2000ApJ...532.1132B, 2006A&A...453..229B}. STELLA is an implicitly differenced hydrodynamics code with a Lagrangian coordinate that incorporates multigroup radiative transfer. It solves the radiative transfer equations in the intensity momentum approximation in each frequency bin at the same time as it solves the equations for mass, momentum, and total energy. 

First, we load the progenitor model at the core infall, excise the core, inject energy until the total energy of the model reaches the specified explosion energy, using MESA. Then, using the same code, we follow the shock propagation until it approaches the shock breakout. These procedures closely follows the \texttt{example\_ccsn\_IIp} test case in MESA. Next, when the shock reaches close to the shock breakout, we use the profile as an input into the public distribution of STELLA included within MESA, and run beyond the plateau phase. At the end of the STELLA run, a post-processing script produces data to compare with observational data.

This time, we did not take into account of the effects of the \citet{2016ApJ...821...76D} prescription for mixing via Rayleigh-Taylor instability. As shown in \citet{2018ApJS..234...34P}, the Rayleigh-Taylor instability slightly affects the light curve and the velocity evolution, especially at around the end of the plateau, when the photosphere passes the H/He boundary. However, the effect on the overall light curve is not so significant.

For the models with the pre-SN energy injection, we assume the explosion energy of $E_\mathrm{exp} = 2.8 \times 10^{51}$ erg. The Ni is given by hand, with the mass assumed to be $M_{\mathrm{Ni}}=0.25M_{\odot}$. This is less than the upper limit of $0.4M_{\odot}$ derived for SN2009kf \citep{2010ApJ...717L..52B}. Besides the model with the initial mass of 12$M_{\odot}$ (inject\_M12), we also calculated the model with the initial mass of 18$M_{\odot}$ (inject\_M18). We mainly focus on the 12$M_{\odot}$ model throughout the paper, while the 18$M_{\odot}$ model is briefly discussed in the section \ref{discussion}.

We have also calculated the models without pre-SN energy injection for the purpose of comparison. For those models without including the pre-SN energy injection, we evolve the progenitors with the initial mass of $12, 18, 20 M_{\odot}$, which covers the possible range for SNe IIP progenitors \citep{2009ARA&A..47...63S, 2015PASA...32...16S}. For each progenitor model , we calculate 3 explosion models. Two of them have the typical values of explosion energy and $^{56}$Ni mass; ($E_{\mathrm{exp}}, M_{\mathrm{Ni}}$) = ($0.5 \times 10^{51}$ erg, 0.03 $M_{\odot}$) and ($1.0 \times 10^{51}$ , 0.03 $M_{\odot}$). The other variant is an energetic explosion with ($E_{\mathrm{exp}}, M_{\mathrm{Ni}}$) = ($4.0 \times 10^{51}$ erg, 0.25 $M_{\odot}$). These values roughly correspond to the high end among typical SNe II \citep{2009ApJ...703.2205K, 2017ApJ...841..127M}. All the models are summarized in the Table \ref{table}.

%\section{result} \label{result}
%% 比較するデータ：Vバンド、bolometricの光度曲線。velocity。

\section{Density profile of the progenitor at the time of the core collapse} \label{density_progenitor}

\begin{figure}%[htbp] % [htbp]
  \hspace{-2.8cm}
  %\vspace{-3cm}
 % \begin{center}
     \includegraphics[width=16cm]{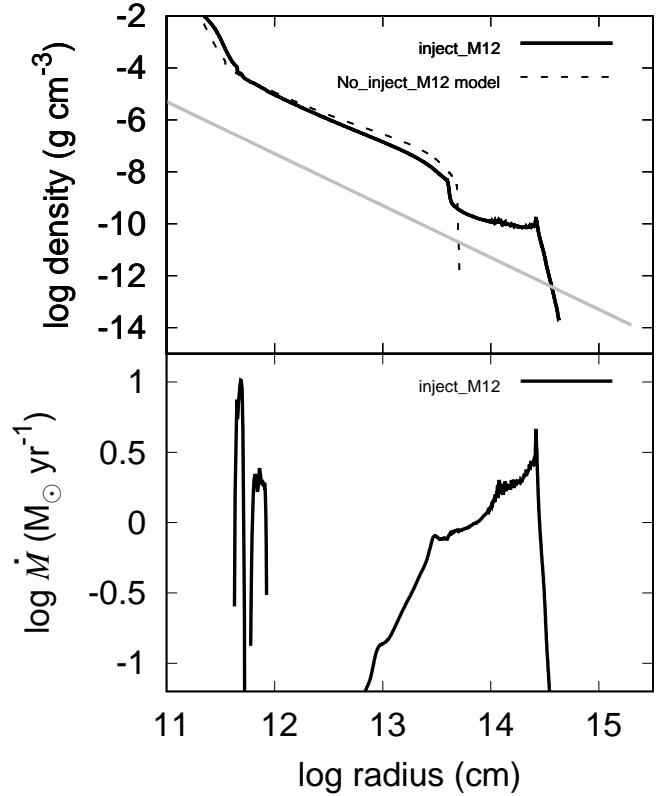}
     \vspace{-0.3cm}
\caption{Top: the density profiles at the time of the core collapse for the $M=12 M_{\odot}$ models. The model with the pre-SN energy injection (inject\_M12) is shown with a black solid line, while a black dashed line shows the density profile of the model without including the pre-SN energy injection (No\_inject\_M12 model). For reference, the density profile for a constant mass flux of $\dot{M} = 10^{-2} M_{\odot} \mathrm{yr}^{-2}$ is also shown with a gray solid line, assuming the constant wind velocity of $v = 10$ km s$^{-1}$. Bottom: the mass flux $\dot{M}=4 \pi r^2 \rho v$ for the model with the pre-SN energy injection (inject\_M12).}
  \label{density_Ldep2.5d39}
%  \end{center}
\end{figure}

Fig.\ref{density_Ldep2.5d39} shows the density and mass flux ($\dot{M} \equiv 4 \pi r^2 \rho v$) profiles at the time of the core collapse for the models with the initial mass of 12$M_{\odot}$ (i.e. inject\_M12 model and No\_inject\_M12 models). For reference, the density profile for a constant mass flux of $\dot{M} = 10^{-2} M_{\odot} \mathrm{yr}^{-1}$ and the constant wind velocity of $v = 10$ km s$^{-1}$, extending to $2 \times 10^{15}$ cm, is also shown. These values are the same as the properties of the CSM suggested by \citet{2011MNRAS.415..199M}.
Since it is not possible to transfer all of the extra heat either by the radiation or convection, the envelope expands significantly, creating the dense wind with the resulting mass flux of $\dot{M} \gtrsim 1 M_{\odot} \mathrm{yr}^{-1}$ above the initial stellar surface. Note that a part of the inner envelope is infalling onto the He core, and this makes the "gap" in the mass-flux profile at around the radius of $\sim 10^{12}$ cm.

 %The wind thus established above the initial surface is close to a steady state in this case, except for the outermost region with the density inversion.
%The reason for this high density shell structure is that the wind velocity increases as time passes due to the lowered density, and the matter launched later catches up with the previously ejected matter, thus making a dense shell \citep{2016MNRAS.458.1214Q}.

\section{Light curves} \label{LC}

\begin{figure*}
 \begin{center}
    \begin{tabular}{c}
    \begin{minipage}{0.5\hsize}
    \begin{center}
      \includegraphics[width=80mm]{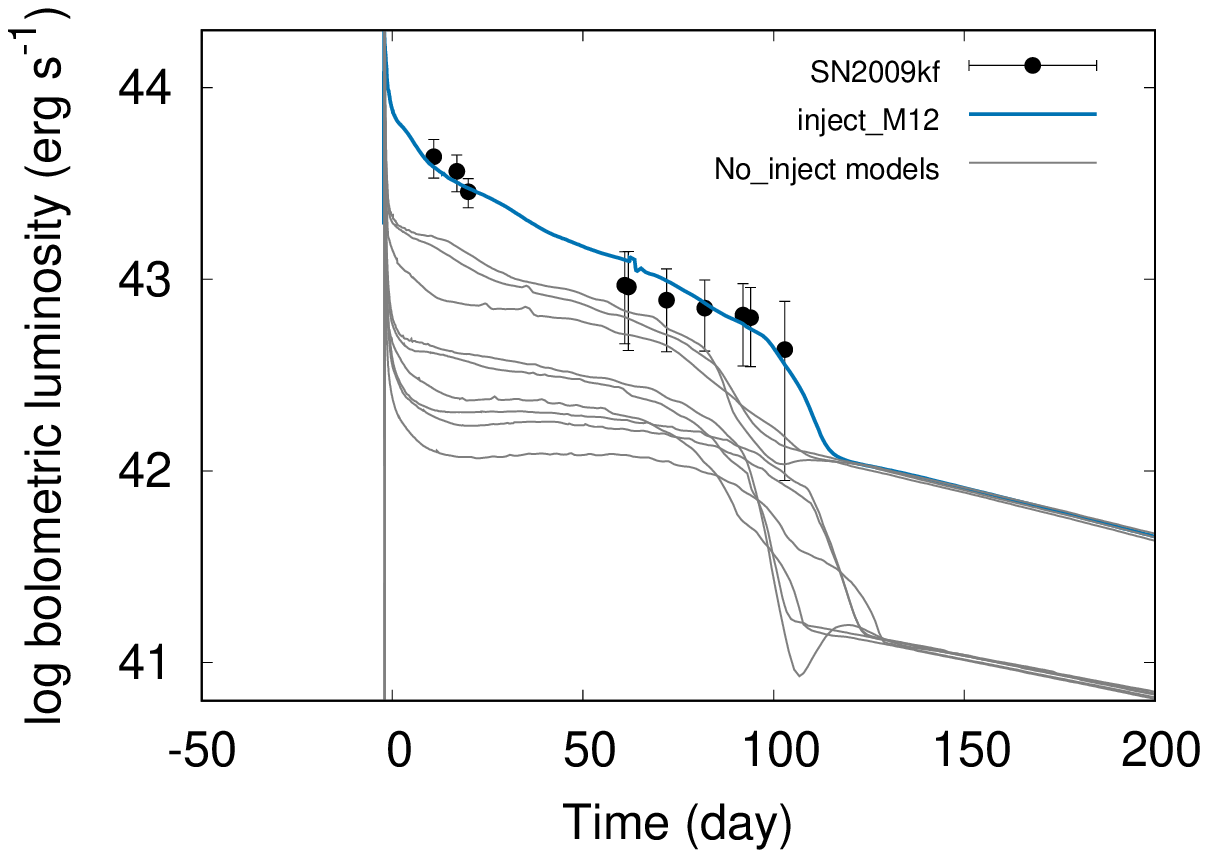}
    \end{center}
  \end{minipage}
  \begin{minipage}{0.5\hsize}
    \begin{center}
       \includegraphics[width=90mm]{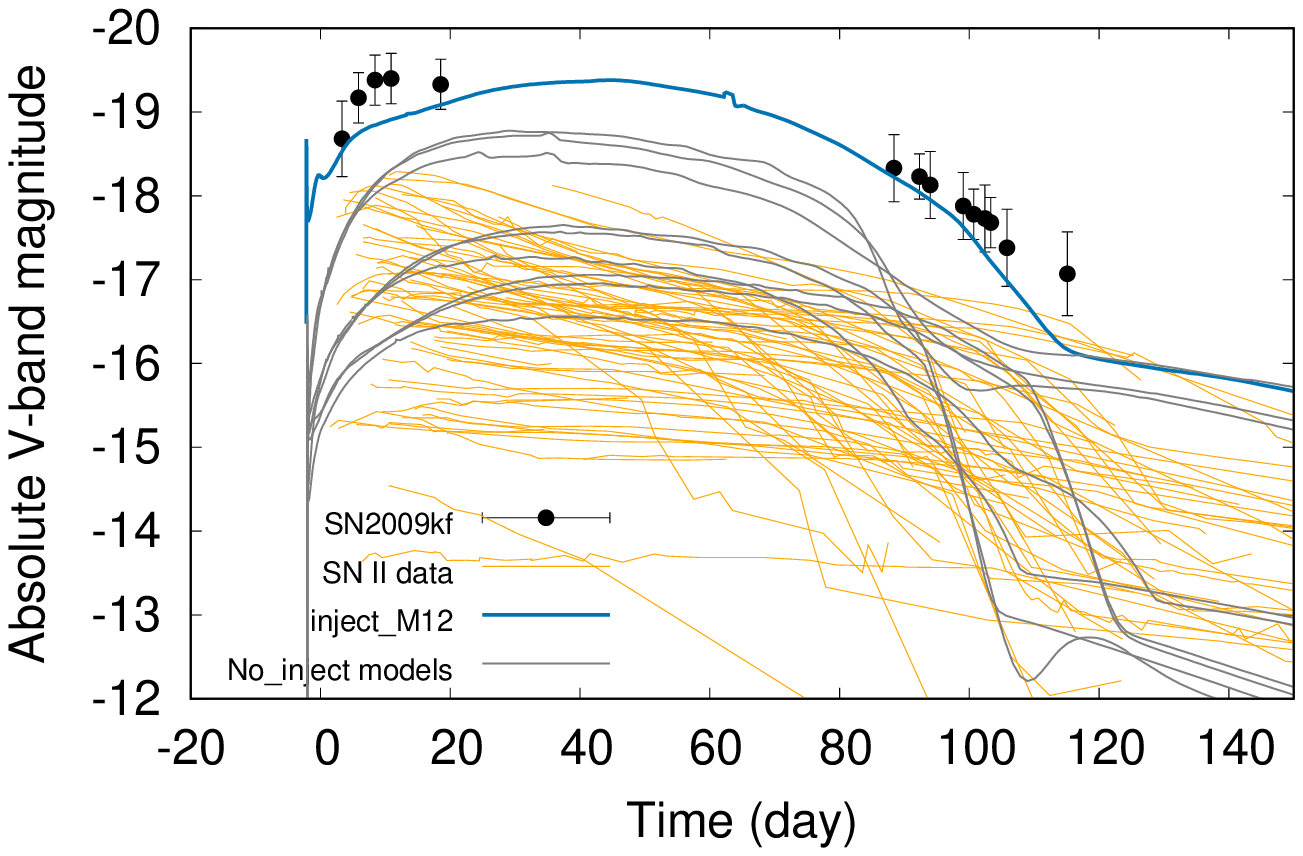}
    \end{center}
  \end{minipage}
  \end{tabular}
 \end{center}
 %\vspace{5zw}
 \caption{Left: The bolometric light curve of our model with the pre-SN energy injection (inject\_M12; blue solid line) compared to the observational data of SN 2009kf taken from \citet{2010ApJ...717L..52B}. Right: The same but for the $V$-band light curve. In both figures, the zero point of the x-axis is the estimated explosion date of SN2009kf \citep{2010ApJ...717L..52B}. All the models are shifted by 2 days relative to the maximum brightness. For reference, the models without the pre-SN energy injection are also shown with black solid lines. In the right panel, the V-band light curves of the observed SNe II are also shown with orange lines \citep{2014ApJ...786...67A}.}
  \label{LCs}
\end{figure*}

Fig.\ref{LCs} compares the bolometric and V-band light curves of our model with the pre-SN energy injection (inject\_M12) to the observational data of SN 2009kf. In both figures, the models without the pre-SN energy injection are also shown. It is evident that the models without the pre-SN energy injection fail to explain the high brightness and long plateau of SN2009kf, even assuming the rather high values of ($E_{\mathrm{exp}}, M_{\mathrm{Ni}}$) = ($4.0 \times 10^{51}$ erg, 0.25 $M_{\odot}$). Thus, we conclude that as long as we consider the typical progenitor and explosion parameters, the models without the pre-SN energy injection never explain the properties of SN2009kf.

On the contrary, our model with the pre-SN energy injection explains both the bolometric and $V$-band light curves well. Especially, our model reproduces explaining the high bolometric luminosity at $\sim$ 15 days and the long plateau. As explained by \citet{2019ApJ...877...92O}, our model behaves like a supernova from a huge red supergiant. Therefore, the plateau becomes more luminous and long, as expected from the scaling relations derived by \citet{1993ApJ...414..712P, 2009ApJ...703.2205K}: $L_{\mathrm{SN}} \propto E^{5/6} M_{\mathrm{ej}}^{-1/2} R_0^{2/3}\kappa^{-1/3}T_{\mathrm{I}}^{4/3}$,  $t_{\mathrm{SN}} \propto E^{-1/6} M_{\mathrm{ej}}^{1/2} R_0^{1/6} \kappa^{1/6} T_{I}^{-2/3}$, where $L_{\mathrm{SN}}$ and $t_{\mathrm{SN}}$ denote the luminosity and duration of the plateau phase. Here, $E$ is the explosion energy, $M_{\mathrm{ej}}$ is the ejecta mass, $R_0$ is the pre-supernova progenitor radius and $\kappa$, $T_{\mathrm{I}}$ are opacity and ionization temperature, respectively.

It is worthwhile to note that \citet{2019ApJ...879....3G} have recently calculated the light curves of SNe IIP for the wide range of parameters using MESA+STELLA. Qualitatively, our models without the pre-SN energy injection show good agreement to their models. Our models, however, show slightly brighter light curves than their models. For example, Fig. 6 in \citet{2019ApJ...879....3G} shows that their model with the parameters of $M_{\mathrm{ZAMS}}=16.3 M_{\odot}$, $R=608 R_{\odot}$, $E_{\mathrm{exp}}=1.0 \times 10^{51}$ erg and $M_{\mathrm{Ni}}=0.03 M_{\odot}$ has the plateau luminosity of log$L \sim42.2$ at 50 days. On the contrary, our models with the same explosion energy and $^{56}$Ni mass have the luminosity (log$L$) of 42.3 and 42.5 for the initial masses of 12$M_{\odot}$ and 18$M_{\odot}$, respectively. Thus, our plateau luminosity is slightly higher than their models. This is due to the larger radii of our progenitor models than theirs (c.f. Table. 1). This larger radii is caused by the smaller mixing length parameters used in our models ($\alpha_{\mathrm{mlt}} = 1.9$) than the one used in their models ($\alpha_{\mathrm{mlt}} = 3.0$) \citep{2018ApJ...853...79C, 2018ApJS..234...34P}.

\section{Blackbody temperature} \label{sec:Tph}

\begin{figure}%[htbp] % [htbp]
  \begin{center}
    \vspace{1.2cm}
     \hspace{-2.0cm}
  \includegraphics[width=9cm, bb = 0 0 370 254]{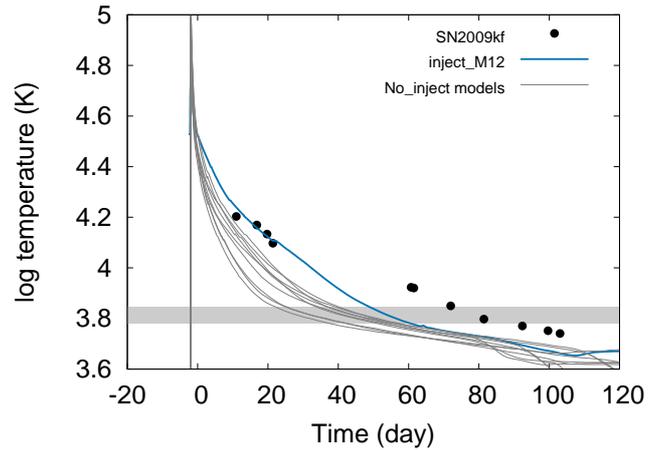}
  \vspace{-1.0cm}
  \caption{The blackbody temperature of our model with the pre-SN energy injection (inject\_M12; blue solid line). For reference, the models without the pre-SN energy injection are also shown with black solid lines. The observational data are taken from \citet{2010ApJ...717L..52B}. The hatched area shows the temperature between $6000$ K and $7000$ K, which corresponds to the hydrogen recombination temperature.}
  \label{fig:Tph}
  \end{center}
%  \vspace{3zw}
\end{figure}

Fig.\ref{fig:Tph} compares the blackbody temperature of our model with the pre-SN energy injection (inject\_M12) to the observational data of SN 2009kf, which are taken from \citet{2010ApJ...717L..52B}. For reference, the models without the pre-SN energy injection are also shown, and the range of typical hydrogen recombination temperature is hatched.

Our model with the pre-SN energy injection explains the data of SN 2009kf reasonably well, especially at the early phase ($t \lesssim 20 $ days). Due to the larger initial progenitor radius, the evolution of the temperature is slower than the models without the pre-SN energy injection, and the recombination of the hydrogen starts only after $t \sim 50$ days.

\section{Photospheric velocity} \label{vph}

\begin{figure}%[htbp] % [htbp]
  \begin{center}
     \hspace{-2.0cm}
  \includegraphics[width=9cm, bb = 0 0 370 254]{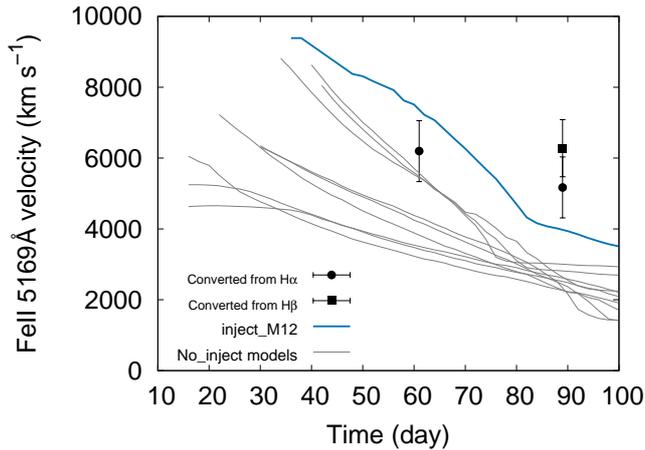}
  \vspace{-1.0cm}
  \caption{The Fe $_{\mathrm{II}} \lambda$ 5169 line velocity of our model with the pre-SN energy injection (inject\_M12; blue solid line). For reference, the models without the pre-SN energy injection are also shown with black solid lines. The observational data of SN2009kf are taken from \citet{2010ApJ...717L..52B}. }
  \label{v_expansion}
  \end{center}
%  \vspace{3zw}
\end{figure}

Between 61 and 89 days after the explosion, SN2009kf showed slow evolution in the Balmer lines. By fitting the Gaussian profiles to these features, the expansion velocity of $9000 \pm 1000$ km s$^{-1}$ is derived from H$\alpha$ line on day 61, and $7800 \pm 1000$ km s$^{-1}$ is derived from H$\alpha$ and H$\beta$ lines on day 89 after the discovery \citep{2010ApJ...717L..52B}.

STELLA outputs the velocity of Fe $_{\mathrm{II}} \lambda$ 5169 ($v_{\mathrm{FeII}}$) line, based on the location where the Sobolev optical depth \citep{1960mes..book.....S} equals unity. This is because the Fe $_{\mathrm{II}} \lambda$ 5169 line is considered to be a good indicator of the photospheric velocity. Note, however, that $v_{\mathrm{FeII}}$ is slightly higher than the photospheric velocity at $\tau = 2/3$ \citep{2019ApJ...879....3G}.

  It is also known that the velocities measured from H$\alpha$ ($v_{H \alpha}$) and H$\beta$ ($v_{H \alpha}$) are generally higher than the velocity measured from Fe $_{\mathrm{II}} \lambda$ 5169 ($v_{\mathrm{FeII}}$) \citep{2012MNRAS.419.2783T, 2014MNRAS.442..844F, 2018A&A...611A..25G}. So, in order to compare $v_{\mathrm{FeII}}$ derived from the STELLA models to the observationally derived velocities of H$\alpha$ and H$\beta$ lines, we need to convert the observationally derived values of H$\alpha$ and H$\beta$ to the Fe $_{\mathrm{II}} \lambda$ 5169 velocities.

\citet{2014MNRAS.442..844F} has derived a linear relation between the velocities measured from the H$\alpha$ line and Fe II $\lambda$ 5169 as $v_{\mathrm{FeII}} = (0.855 \pm 0.006)v_{\mathrm{H} \alpha} - (1499 \pm 87)$ km s$^{-1}$. They also derived the linear relation of $v_{\mathrm{FeII}} = (0.805 \pm 0.005)v_{\mathrm{H} \beta}$ between $v_{H \beta}$ and $v_{\mathrm{FeII}}$. Here, we use these relations to convert $v_{H \alpha}$ and $v_{H \beta}$ which are taken from \citet{2010ApJ...717L..52B} into $v_{\mathrm{FeII}}$ so that we can compare them to the STELLA models. Note, however, that these relations are derived for typical SNe IIP. These relations may not be applicable to SN2009kf, which is a quite peculiar object.
% Fang et al. をチェック。relationはどういう天体に対して得られている？

Fig. \ref{v_expansion} compares $v_{\mathrm{FeII}}$ of our STELLA models to $v_{\mathrm{FeII}}$ observationally derived for SN2009kf. Our model with the pre-SN energy injection (inject\_M12) fits the velocity of SN2009kf reasonably well. Although the fit at 89 days is not very good, we do not take this as a strong tension; the linear relations between $v_{H \alpha}$, $v_{H \beta}$ and $v_{\mathrm{FeII}}$ may not be applicable to the peculiar SNIIP SN2009kf and such relations are known to be time-dependent \citep{2018A&A...611A..25G}. Note also that the velocity evolution at around the end of plateau phase is affected by the Rayleigh-Taylor instability, which we did not consider in the present work. Actually, \citet{2018ApJS..234...34P} show that the velocity drops slowly at the plateau phase, if they include the prescriptions of the Rayleigh-Taylor instabilities. Thus, such an effect may help to improve the fit especially in the late phase. %Thus, given the uncertainty in deriving the Fe II $\lambda$ 5169 velocity, this comparison should be regarded only as being indicative.

\section{Discussion} \label{discussion}
Assuming there is a strong pre-SN energy injection to the envelope ($L_{\mathrm{dep}}= 3.0 \times 10^{39}$ erg s$^{-1}$) in the last few years of the stellar evolution, we have shown that the observational properties of SN2009kf are naturally explained with an explosion energy in the range seen in other SNe IIP. The explosion energy we have thus derived is $E_{\mathrm{exp}} = 2.8 \times 10^{51}$ erg s$^{-1}$, which is an order of magnitude smaller than the previous estimate by \citet{2010ApJ...723L..89U}. Although this value is still high, it is still within the range of the explosion energy derived for SNe IIP from observations \citep{2003ApJ...582..905H, 2011ApJ...729...61B} %, 2011MNRAS.417..261I}
%, 2013A&A...555A.142I}.

\begin{figure}%[htbp] % [htbp]
    \begin{center}
      \hspace{-2.0cm}
   %   \vspace{3.3cm}
  \includegraphics[width=9cm, bb = 0 0 370 254]{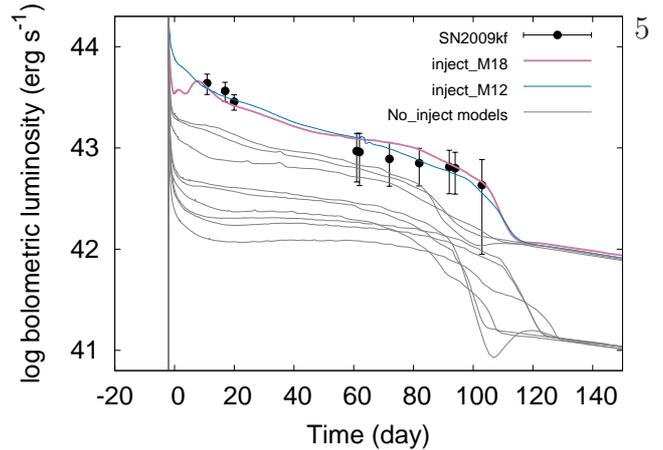}
  \vspace{-1.0cm}
  \caption{The bolometric light curve of our model with the initial mass of $M=18M_{\odot}$, including the pre-SN energy injection (inject\_M18; pink solid line). For reference, the model inject\_M12 (blue solid line) and the models without the pre-SN energy injection (black solid lines) are also shown. The observational data are taken from \citet{2010ApJ...717L..52B}. }
  \label{Lbol_M18}
  \end{center}
%  \vspace{3zw}
\end{figure}

% Ni mass についてもコメント。

%CSMの質量に関してmoriyaと比較。<=いらないかな？？
%%\citet{2011MNRAS.415..199M} have demonstrated that the observational properties of SN2009kf is naturally explained by attaching a dense CSM above the stellar surface. They derived the mass-loss rate of $\dot{M} = 10^{-2} M_{\odot}$ yr$^{-1}$, which extends to $2 \times 10^{15}$ cm (gray line in Fig.\ref{density_progenitor}). As is clear from Fig.\ref{density_progenitor}, the density of our progenitor model with pre-SNenergy injection is much higher, and the radial extent of our model is smaller than the model by \citet{2011MNRAS.415..199M}. It is known, however, that there is a degeneracy between these two parameters (the density and radial extent of the CSM). Actually, \citet{2017ApJ...838...28M} have shown that the models with extended low-density wind produce similar light curves to the models with less extended high-density wind. In this sense, it is true that our model is similary to the model of \citet{2011MNRAS.415..199M}, but we have shown that the \textcolor{red}{dense CSM} is the natural consequence of the energy injection to the envelope in the last few years of stellar evolution.

% waveの観点で言えること。

With the injection rate of $L_{\mathrm{dep}} =3.0 \times 10^{39}$ erg s$^{-1}$ and the injection duration of $3.0$ years, the total extra energy that is injected is $\sim 3 \times 10^{47}$ erg. In the present study, we did not consider any specific mechanism for this energy injection.　Note, however, that the total deposited energy is consistent with the value available by the waves. \citet{2014ApJ...780...96S} have estimated the wave energy excited during the advanced burning stages. In their model of $M_{\mathrm{ZAMS}} =  12M_{\odot}$, $Z=Z_{\odot}$, which assumes no rotation, the total wave energy excited during the core neon burning, core oxygen burning, and core silicon burning amounts to $\approx 3 \times 10^{47}$ erg, which is consistent with the value in our model \citep[see Table 2 in][]{2014ApJ...780...96S}. 

We have focused on the model with the initial mass of $12M_{\odot}$. We are, however, not insisting that this is a unique solution for SN2009kf. In Fig. \ref{Lbol_M18}, we compare the pre-SN model with the initial mass of 18$M_{\odot}$ (inject\_M18) to the data of SN2009kf. The parameters are set as $L_{\mathrm{dep}}= 2.0 \times 10^{39}$ erg s$^{-1}$ and $R_{\mathrm{dep}}=940 R_{\odot}$ in this model (see Table1). As is shown in Fig. \ref{Lbol_M18}, this model also explains the bolometric light curves of SN2009kf. This implies that a series of models with different progenitor masses can explain the light curve of SN2009kf, given that the rate of the pre-SN energy injection is set smaller for a more massive progenitor. This shows that our argument for the pre-SN energy injection as a possible origin of bright SNe IIP is not dependent on any particular model parameters. We note that our aim in this work is to show the applicability of such a scenario rather than trying to find a unique set of model parameters.

As another possible candidate for the scenario, SN IIP LSQ13fn are known to have high luminosity and low velocity, compared with typical SNe IIP \citep{2016A&A...588A...1P, 2020MNRAS.494.5882R}. With the higher energy injection rate or the longer duration of the energy injection than what we considered in the present work, the resulting SN should have higher luminosity and lower velocity \citep{2019ApJ...877...92O}. Therefore, our models with the different parameters might be applicable to such objects, which we would like to investigate in the future.

% high-L & low-velが説明できるかも。

\section*{Acknowledgements}
We thank the anonymous referee for the constructive comments. We thank Maria Teresa Botticella for providing us with the observational data of SN 2009kf. We also thank Sung-Chul Yoon for useful discussion. R. O. acknowledges support from JSPS Kakenhi grants (19J14158). K. M. acknowledges support from JSPS Kakenhi grants (18H05223, 18H04585 and 17H02864).

\section*{Data availability}
The data underlying this article were provided by Maria Teresa Botticella by permission. Data will be shared on request to the corresponding author with permission of Maria Teresa Botticella.



%%%%%%%%%%%%%%%%%%%%%%%%%%%%%%%%%%%%%%%%%%%%%%%%%%



%\bibliographystyle{mnras}

%%%%%%%%%%%%%%%%%%%%%%%%%%%%%%%%%%%%%%%%%%%%%%%%%%

%%%%%%%%%%%%%%%%% APPENDICES %%%%%%%%%%%%%%%%%%%%%

%\appendix



%%%%%%%%%%%%%%%%%%%%%%%%%%%%%%%%%%%%%%%%%%%%%%%%%%

% Don't change these lines
\bsp	% typesetting comment
\label{lastpage}
\end{document}